\documentclass{article}[12pt]
\newcommand{\eqref}[1]{(\ref{#1})}

\def\cosech{{\rm cosech\,}}
\def\coth{{\rm coth\,}}
\def\sinh{{\rm sinh\,}}
\def\sech{{\rm sech\,}}

\def\sn{{\rm sn\,}}
\def\cn{{\rm cn\,}}
\def\dn{{\rm dn\,}}

\usepackage{array}
\usepackage{graphicx,subfigure}
\usepackage{epsfig}
\begin{document}
\begin{center}
{\bf \LARGE {Soliton response to transient trap variations }}\\
\vspace{0.5cm}
S. Sree Ranjani$^\dag$\footnote{ssrsp@uohyd.ernet.in}, Utpal Roy$^\ddag$\footnote{utpal.roy@unicam.it}, P. K. Panigrahi$^\ast$\footnote{prasanta@prl.res.in} and A. K. Kapoor$^\dag$\footnote{akksp@uohyd.ernet.in},\\
$^\dag$ School of Physics, University of Hyderabad, Hyderabad 500 061, India. \\
$^\ddag$ Dipartimento di Fisica, Universit$\grave{a}$  di Camerino, I - 62032 Camerino, Italy.\\
$^\ast$  Indian Institute of Science Education and Research (IISER) Kolkata, Salt Lake, Kolkata 700 106, India.\\
\end{center}

\begin{abstract}
The response of bright and dark solitons to rapid variations in an expulsive longitudinal trap is investigated. We concentrate on the effect of transient changes in the trap frequency in the form of temporal delta kicks and the hyperbolic cotangent functions.
Exact expressions are obtained for the soliton profiles. This is accomplished using the fact that a suitable linear Schr\"odinger
stationary state solution in time can be effectively combined with the solutions of non-linear Schr\"odinger equation, for
obtaining solutions of the Gross-Pitaevskii equation with time dependent scattering length in a harmonic trap. Interestingly, there is rapid pulse amplification in certain scenarios.
\end{abstract}

{\bf PACS} number(s) : 03.75-b, 05.45.Yv\\

\noindent
{\bf I Introduction}\\

Bose-Einstein condensates (BECs) \cite{bose} - \cite{petB} can be confined  to one and two dimensions in suitable traps \cite{bag} - \cite{ket}, which has opened up the possibility for their technological applications. BEC on  a chip is a two dimensional configuration \cite{pro}, whereas cigar shaped BECs refer to quasi one dimensional scenario \cite{jack} - \cite{kam}. In the latter case, the transverse confinement is taken to be strong, with the longitudinal trap frequency having comparatively much smaller value. In this condition, dark \cite{ber2} - \cite{ber1} and bright \cite{ADJ} - \cite{cor} solitons have been observed in repulsive and attractive coupling regimes respectively. In the context of the BEC, bright and dark solitons respectively correspond to density lumps and  rarefactions compared to the background.  The corresponding mean field Gross-Pitaevskii (GP) equation being the familiar non-linear Schr\"odinger equation (NLSE) in quasi-one dimensions \cite{zha}, the observation of localized solitons has demonstrated the ubiquity of solitons in diverse branches of physics, which share common non-linear behavior.

Very interestingly, in the cigar shaped BEC, a number of parameters like the scattering length and trap frequencies can be changed in a controlled manner, leading to the possibility of coherent control of solitons.  The former can have a temporal variation due to Feshbach resonance, a scenario which has been extensively studied in the literature \cite{tie} - \cite{dic}. In particular, the discovery of bright soliton and soliton trains has led to a systematic investigation of the impact of the change in the scattering length on the BEC profile. Recently, the temporal modulation of the non-linearity in one dimension arising due to the  corresponding variations in the transverse trap frequency have been systematically studied in the context of generation of Faraday modes \cite{kes}, \cite{eng}.

In a recent paper \cite{atre}, a method to obtain analytical solutions of the GP equation for a range of temporal variations of parameters like the scattering length, the longitudinal trap frequency etc. was demonstrated.   It was shown that in the process of reducing the GP equation to a known non-linear equation, having Jacobi elliptic functions as solutions, the consistency conditions obtained mapped onto the linear Schr\"odinger equation. This mapping enables us to obtain exact solutions of the GP equation for a wide range of temporal modulations of the control parameters. These solutions exhibit soliton trains, localized solitons (both bright and dark) for different regimes of the coupling parameter. Interestingly, exact solutions of the GP equation have been obtained for other kind of trap potentials, for example linear traps \cite{alk} using Darboux transformations. The effect of  the variations in the time dependent coupling on the soliton profiles has also received a lot of attention \cite{alk} - \cite{rad}.

The goal of the present paper is to investigate the soliton dynamics in a BEC in an expulsive harmonic trap, whose frequency is time dependent. The expulsive trap accelerates the soliton. Very interestingly it is found that for certain conditions it can be compressed, particularly under a sudden transient variation in the trap frequency. As we have mentioned earlier, a temporal modulation in the transverse frequency leads to the generation of Faraday modes in the longitudinal direction. Hence, it is of interest to study the effect of transient variations in the longitudinal trap on the BEC profiles. We concentrate here on transient variations and for explicitness we use the delta functions and the hyperbolic cotangent functions to model rapid temporal variations in the trap frequency.  We obtain exact solutions and study the  spatio-temporal dynamics of the solitons and their response to the changes in the trap frequency. We point out that harmonic traps decorated with delta functions and a dimple potential have been recently studied numerically  \cite{un} and the properties like atomic density, chemical potential, critical temperature etc. of the condensate have been obtained.

 In the following section, we obtain the exact solutions of the GP equation and illustrate how we can modulate the expulsive trap frequency using suitable quantum mechanical models \cite{atre}. In section III, the trap frequency is transiently changed by hyperbolic cotangent functions. In section IV, we study the response of the solitons to sudden changes in the trap frequency in the form of delta kicks. For all the cases studied we obtain profiles of both dark and bright soliton trains and localized soliton solutions. The spatio-temporal dynamics of the soliton solutions and the behavior of the non-linearity with time in the attractive and repulsive coupling regimes are plotted. It is found that for certain examples there is an amplification in the soliton amplitude resulting in its compression while in other examples the solitons spread with a decaying amplitude. \\

\noindent
{\bf II The Dynamics of the Bose-Einstein condensate}\\

   For a BEC at zero temperature, confined in  cylindrical harmonic trap $V_0(x,y)=m^{\prime}\omega^2_{\bot}(x^2+y^2)/2$ and  time dependent harmonic confinement along the $z$ direction $V_1(z,t) =m^{\prime}\omega^2_0(t)z^2/2$, the dynamics is governed by the GP equation
\begin{equation}
i\hbar \frac{\partial \Psi({\bf r},t)}{\partial t} = \left(-\frac{\hbar^2}{2m^{\prime}}\nabla^2 +U|\Psi({\bf r},t)|^2 +V\right)\Psi({\bf r},t),  \label{e1a}
\end{equation}
where $V=V_0+V_1$, $m^{\prime}$ is the mass of the particle and $U=4\pi\hbar^2a_s(t)/m^{\prime}$ gives the non-linear coupling with $a_s(t)$ denoting the scattering length. The nonlinearity can be either attractive $(a_s <0)$ or repulsive $(a_s>0)$. In order to reduce the $3D$ GP equation to quasi-one dimensions, it is assumed that the interaction energy of atoms is much less than the kinetic energy in the transverse direction \cite{sal}.
Substituting the following trial wave function in dimensionless units,
\begin{equation}
\Psi({\bf r},t) =\frac{1}{\sqrt{2\pi a_B a\bot}} \psi\left(\frac{z}{a_\bot},\omega_\bot t\right)\exp\left(-i\omega_\bot t -\frac{x^2+y^2}{2a^2_\bot}\right),    \label{e1b}
\end{equation}
we obtain the quasi-1D NLSE
\begin{equation}
\imath\partial_{t}\psi=-\frac{1}{2}\partial_{zz}\psi+\gamma(t)|\psi|^{2}\psi+\frac{1}{2}M(t)z^2\psi.    \label{e1}
\end{equation}
Here, $\gamma(t)= 2a_s(t)/a_B$, $M(t)=\omega_0^2(t)/\omega^2_{\perp}$,
$a_{\perp}=(\hbar/m^{\prime}\omega_{\perp})^{1/2}$ with $a_B$ being the Bohr's radius. We have kept $M(t)$, related to the trap frequency, time dependent. This enables us to modulate the trap frequency, allowing us to study the effect of different temporal profiles of the trapping potential on the solitons. The specific case of $M(t)$ being a constant, describes the scenario where the oscillator potential is either confining $(M > 0)$ or repulsive $(M <0)$. Analytic solutions to the above NLSE are obtained using the ansatz solution
\begin{equation}
\psi(z,t)=\sqrt{A(t)}F\{A(t)[z-l(t)]\}\exp[\imath\Phi(z,t)],
\label{e2}
\end{equation}
where $A(t)$ describes the amplitude and $l(t) =\int_0^t v(t^{\prime})dt^{\prime}$ determines the location of the center of mass. We assume the phase to be of the form
\begin{equation}
\Phi(z,t)= a(t)-\frac{1}{2}c(t)z^2. \label{e3}
\end{equation}
Substitution of $\psi(z,t)$ in the NLSE gives us the consistency conditions
\begin{equation}
A(t)=A_0\exp \left(\int_0^t c(t^\prime)dt^{\prime}\right),\,
\frac{dl(t)}{dt} +c(t)l(t)=0  \label{e6a}
\end{equation}
and
\begin{equation}
\gamma(t)=\gamma_0 \frac{A(t)}{A_0},    \label{e6b}
\end{equation}
where $A_0, \gamma_0$ and $l_0$ are constants. We also obtain $a(t)=a_0+\frac{\lambda -1}{2}\int_0^t A^2(t^{\prime})dt^{\prime}$ along with the Riccati
equation for $c(t)$,
\begin{equation}
\frac{dc(t)}{dt}-c^2(t)=M(t).  \label{e4}
\end{equation}
From the above equations, it can be seen that $A(t),\gamma(t)$ and $l(t)$ are all non-trivially related to the phase component $c(t)$. Added to this, the fact that the Riccati equation can be mapped onto the linear Schr\"odinger eigenvalue problem with potential $V(t)$,
\begin{equation}
-\phi^{\prime\prime}(t)-M(t)\phi(t)=0 \label{e6}
\end{equation}
with $M(t)= M_0-V(t)$, using the change of variable,
\begin{equation}
c(t)=-\frac{d}{dt}\ln[\phi(t)] \label{e5}
\end{equation}
gives us control over the dynamics of the BEC.  Hence, knowing $c(t)$ or the solution $\phi(t)$ of \eqref{e6}, allows us to obtain analytical expressions of the control parameters through equations \eqref{e5} and \eqref{e6a}-\eqref{e6b}. It is a well known fact that for many potential models, the Schr\"odinger equation can be solved exactly. In the next sections, we exploit this  and the connection between $c(t)$ and $\phi(t)$  to study the dynamics of the BEC analytically, when the oscillator frequency undergoes a rapid change.

   The use of the above consistency conditions along with the ansatz solution in the NLSE gives us the following differential equation for $F$, in terms of the new variable $T=A(t)[z-l(t)]$,
\begin{equation}
F^{\prime\prime}(T)-\lambda F(T) +2\kappa F^3(T)=0  \label{e3a}
\end{equation}
with $\kappa=-\frac{\gamma_0}{A_0}$ and the differentiation is with respect to $T$. The general solution of this differential equation can be written in terms of the $12$ Jacobi elliptic functions $\sn(T,m)\,,\,\cn(T,m) \,,\, \dn(T,m)$ etc where $m$ is the elliptic modulus \cite{han}. The solitons will arise in the limit $m\rightarrow 1$, where $\dn(T,m) = \cn(T,m) \rightarrow \sech(T)$ and $\sn(T,m) \rightarrow \tanh(T)$.

Thus for $\gamma_0 < 0$, bright soliton trains of the form
\begin{equation}
\psi(z,t)=\sqrt{A(t)}\,\,\cn(T/\tau_0,m)\exp[i\Phi(z,t)] \label{e11}
\end{equation}
exist, where $\tau^2_0=-m A_0/\gamma_0$ and $\lambda=(2m-1)/\tau^2_0$ (these coefficients are obtained using equations \eqref{e6a} -\eqref{e6}. In the limiting case $m=1$, equation \eqref{e11} corresponds to a bright soliton solution. It has been seen that for $\gamma_0 >0$, there exist dark soliton trains of the form
\begin{equation}
\psi(z,t)=\sqrt{A(t)}\,\,sn(T/\tau_0,m)\exp[i\Phi(z,t)],
\label{e11a}
\end{equation}
with $\tau^2_0=m A_0/\gamma_0$, $\lambda=-(m+1)/\tau^2_0$ and in the limit $m\rightarrow 1$, the above solution corresponds to a dark soliton.

  In the following sections, we obtain the profiles of bright and dark soliton trains and solitons in a BEC in an expulsive oscillator potential, with experimentally achievable  temporal variations  of the trap frequency. As seen in \eqref{e4}, the oscillator frequency $M(t)$ depends on $c(t)$.   We first concentrate on the response of the soliton profiles to the variations in the trap frequency when $c(t)$ is equal to the hyperbolic cotangent functions.\\

\noindent
{\bf III  Dynamics of the BEC with ${\bf c(t)=\pm \,B\, \coth(t)}$}\\

In this section, we look at the examples where the trap frequency of the confining trap is modulated using $c(t)=\pm \,B \coth(t)\,,\,\, B>0$. Using the consistency conditions in \eqref{e6a} and \eqref{e6b} we obtain the analytical solutions of the NLSE as shown below.\\

\noindent
{\bf case (a)}\\
We take $c(t)=  B\coth(t),\,\, B>0$ in \eqref{e4} and obtain
\begin{equation}
M(t)= -B^2 - B(B+1)\cosech^2(t).             \label{e24}
\end{equation}
Since $M_0=-B^2$, we are looking at an expulsive oscillator scenario. We point out that $V(x)=B(B+1)\cosech^2(x)$ corresponds to the supersymmetric Rosen-Morse potential with superpotential $W=B \,\coth(x)$ \cite{khare}. The function $V(t)$ has a singularity at $t=0$ which causes a sudden change in the trap frequency. Substituting for $c(t)$ in equations \eqref{e6a} and \eqref{e6b} one obtains
\begin{equation}
 a(t)= a_0 - \frac{(\lambda -1)A^2_0}{2}\left(\sinh(t)-4t \right) \label{e25}
\end{equation}
\begin{equation}
A(t)=A_0 \left(\frac{\sinh(t)}{\sinh(t_0)}\right)^B\,,\,  \gamma(t)= \gamma_0 \left(\frac{\sinh(t)}{\sinh(t_0)}\right)^B\,,\, l(t) = l_0 \left(\frac{\sinh(t_0)}{\sinh(t)}\right)^B, \label{e26}
\end{equation}
which when substituted in \eqref{e11} and \eqref{e11a}, gives the following solutions describing the bright and dark soliton trains
\noindent
\begin{equation}
\psi(z,t)=\sqrt{A_0 \sinh(t)}\,\cn\left(A_0 \sinh(t)\frac{[z-l_0 \cosech(t)]}{\tau_0},m\right)\exp(i\Phi(z,t)),  \label{e23a}
\end{equation}
\begin{equation}
\psi(z,t)=\sqrt{A_0 \sinh(t)}\,\sn\left(A_0 \sinh(t)\frac{[z-l_0 \cosech(t)]}{\tau_0},m\right)\exp(i\Phi(z,t))  \label{e23b}
\end{equation}
respectively. The dynamics of these solutions are plotted in fig 1(a) and 1(b), where the matter wave density $|\psi(z,t)|^2$ is plotted for varying $t$ and $z$. We can see that the soliton trains get amplified with time as we move away from the origin.
In the limit $m\rightarrow 1$, from \eqref{e23a} and \eqref{e23b} we get the bright and dark solitons, plotted in fig 1(c) and 1(d) respectively. These solitons are highly localized and are seen to be diverging from the $t=0$ line in the $t-z$ plane. By changing the initial time to some $t_0$, instead of $t=0$ in the consistency conditions and varying $l_0$, the location of the center of mass can be changed. The non-linearity $\gamma(t)$ is plotted against varying $t$, for the attractive $(\gamma_0 <0)$  and the repulsive   $(\gamma_0 > 0)$ coupling regimes in fig 1(e) and 1(f) respectively and we see that for a given $\gamma_0$, the non-linearity does not change sign implying we are either in an attractive or repulsive coupling regimes. By changing the constants $B$, $A_0$, $\gamma_0$, $l_0$ and $t_0$ the amplification, location of the solitons can be changed.  \\

\noindent
{\bf Case (b)}\\
For $c(t)= - B\,\coth(t), \,\,\, B>0$, $M(t)= -B^2-B(B-1)\cosech^2(t)$. Here again we are looking at the expulsive oscillator scenario and the potential $B(B-1)\cosech^2(x)$ corresponds to the supersymmetric Pos\"chl-Teller potential \cite{khare}.  Moreover, this case is interesting because for $B=1$, this potential corresponds to the free particle problem which has been studied in \cite{atre}. Substituting $c(t)$ in equations \eqref{e6a} and \eqref{e6b}, we obtain
\begin{equation}
A(t)=A_0 \left(\frac{\sinh(t_0)}{\sinh(t)}\right)^B\,,\,  \gamma(t)= \gamma_0 \left(\frac{\sinh(t_0)}{\sinh(t)}\right)^B\,,\, l(t) = l_0 \left(\frac{\sinh(t)}{\sinh(t_0)}\right)^B. \label{e23}
\end{equation}
As in the above case, we obtain the profiles of both bright and dark soliton trains and localized solitons which are plotted in figure 2(a) - 2(f). Although the function $V(t)$ has a singularity at $t=0$, like in case (a), the response of the soliton solutions is entirely different. We see that the soliton trains have a very high amplitude around $t=0$ and they decay as we move away from the origin i.e. the soliton profile spreads with time. The amplitude and the decay time can be controlled by changing $A_0$. The solitons plotted in figures 2(c) and 2(d) are again highly localized, but in this case converge towards the $t=0$ line in the $t-z$ plane. From figures 2(e) and 2(f), we see that though the non-linearity does not change sign for a given $\gamma_0$, it does increase rapidly in magnitude around $t=0$, unlike the previous case were $\gamma(t)$ varied slowly with time.\\

\noindent
{\bf IV Response of BEC to delta kicks in the trap frequency }\\

\noindent
In this example, we vary the trap frequency with an attractive delta potential located at the origin and see how the solitons respond to a sudden kick.  Here,
\begin{equation}
V(t)= -V_0\, \delta(t),\,\,\, V_0 > 0.             \label{e17}
\end{equation}
and the solutions of \eqref{e6} are
\begin{eqnarray}
\phi_1(t) &= &\sqrt{\frac{V_0}{2}}\exp(-\frac{V_0}{2}t)  \,\,\,\,\, (t>0) \nonumber \\
        &=&\sqrt{\frac{V_0}{2}}\exp(\frac{V_0}{2}t)    \,\,\,\,\,\,\,\,\,  (t<0)  \label{e18}
\end{eqnarray}
\begin{eqnarray}
\phi_2(t)&=&\sqrt{\frac{V_0}{2}}\exp(\frac{V_0}{2}t) \,\,\,\,\,\,\,\,\,\,\, \,\,\,\,\,\,(t>0) \nonumber \\
        &=&-\sqrt{\frac{V_0}{2}}\exp(-\frac{V_0}{2}t)    \,\,\,\,\,\,\,\,  (t<0)      \label{e18a}
\end{eqnarray}
with $M_0 =-\frac{V^2_0}{2}$. In order to obtain $\phi_2(t)$, we have made use of the fact that the second solution of a second order differential equation can be obtained from the first solution using the relation $\phi_2(t)=\phi_1(t)\int^t[dt^{\prime}(\phi_1(t^{\prime}))^{-2}]$. We can write $c(t)$ in terms of $\phi(t)$ using \eqref{e5} and obtain the consistency conditions in terms of $\phi(t)$ as
\begin{eqnarray}
A(t)=A_0\frac{\phi(0)}{\phi(t)}
\,\,;\,\,\gamma(t)=\gamma_0\frac{\phi(0)}{\phi(t)}\,,\,\,l(t)=l_0\frac{\phi(t)}{\phi(0)}.
\label{e7}
\end{eqnarray}
Now we proceed to study the soliton profiles using the two solutions given in \eqref{e18} and \eqref{e18a} separately.\\

\noindent
{\bf Case (a)}\\
First we consider the solutions in \eqref{e18} and from equation \eqref{e5} we obtain $c(t)= V_0/2$ for $t>0$ and $c(t)= -V_0/2$ for $t<0$. From   \eqref{e7}, we obtain
\begin{equation}
A(t)=A_0 \exp(\frac{V_0}{2}|t|)\,,\,  \gamma(t)= \gamma_0  \exp(\frac{V_0}{2}|t|)\,,\, l(t) = l_0 \exp(-\frac{V_0}{2}|t|) \label{e19}
\end{equation}
and also find that
\begin{equation}
a(t)= a_0 + \frac{(\lambda -1)A^2_0}{2V_0}\left[ \exp(V_0|t|) -1 \right]. \label{e18b}
\end{equation}

Substituting $A(t), \gamma(t)$ and $l(t)$ in \eqref{e11}, we obtain the solutions describing
bright and dark soliton profiles, which are plotted in figure 3 along with the plots showing the behavior of the nonlinearity. From the figures we can see that the sudden temporal kick to the soliton profile results in rapid amplification of the soliton trains over time. Similar to the case (a) of the previous section, the solitons are highly localized and diverge from the origin. By varying $V_0$, we can create a soliton with the required amplitude. The non-linearity $\gamma(t)$ for this case turns out to be an exponential function. Thus the scattering length $a_s(t)$ turns out to be an exponential in time \cite{lia}.  It is known that the dynamics of a BEC in an expulsive parabolic potential are governed by an NLSE with an exponentially varying scattering length. The conditions in which these systems are one dimensional and  the range of the parameter values for which the condensate is stable  are discussed in \cite{kha}.   \\

\noindent
{\bf Case (b)}\\
 Here, we repeat the above analysis with the solution given in \eqref{e18a}. For this solution, the nonlinearity $\gamma(t)$ changes sign when we go from the negative time domain to the positive time domain. Here, we study the soliton dynamics for $t>0$ and the corresponding expressions are
 \begin{equation}
c(t) = -\frac{V_0}{2}\,, \,\, a(t)= a_0 - \frac{(\lambda -1)A^2_0}{2V_0} \left[ \exp(- V_0 t) - 1 \right], \label{e20a}
\end{equation}
\begin{equation}
A(t)=A_0 \exp(-\frac{V_0}{2}t)\,,\,  \gamma(t)= \gamma_0  \exp(-\frac{V_0}{2}t)\,,\, l(t) = l_0 \exp(-\frac{V_0}{2}t). \label{e20b}
\end{equation}
The soliton profiles are plotted in figure 4. We observe that the soliton train amplitude decays rapidly as $t$ increases unlike the previous case. For smaller values of $V_0$, the trains exist for longer time. Thus the decay of the solutions can be controlled by varying $V_0$.\\

In conclusion, we have analytically treated the effect of variations of the longitudinal trap frequency on the BEC profile.  We have looked at the cases where the trap frequency is modulated using hyperbolic cotangent functions and delta kicks. It is observed that in the gain/loss less scenario, these variations in the trap frequency resulted in the soliton profiles either amplifying or spreading. We have also shown that we can manipulate the amplification and the location of the solitons by varying the control parameters.

The fact that a stationary Schr\"odinger eigenvalue problem can be combined with solutions of non-linear Schr\"odinger equation to obtain the solutions of Gross-Pitaevskii equation with time dependent parameters, enabled us to obtain the exact dynamical behavior. As mentioned in the introduction, there are other methods of obtaining exact solutions of the GP equation. The method used in this paper is simpler and the fact that the Schr\"odinger equation can be solved exactly for many potential models, offers us a huge choice of temporal variations of the trap frequency from the class of exactly solvable models. We can choose the models which can be realized experimentally. Along with this wide choice,  the other advantage is that we have total control over the soliton dynamics. It would be interesting to see how solitons evolve in space and time for other choices of variations in the trap frequency. Another interesting problem will be the study of two soliton dynamics under similar variations in the trap frequency \cite{rad}. \\

\noindent
{\bf Acknowledgments}\\
S. S. R. acknowledges financial support provided by Council of Scientific and Industrial
Research (CSIR), Government of India.
\\

\noindent
{\bf References}\\

\newpage
\begin{figure*}
\centering \subfigure[Bright soliton train, $m=0.5$ ,
$\tau_0=0.375$]
{\includegraphics[scale=0.4]{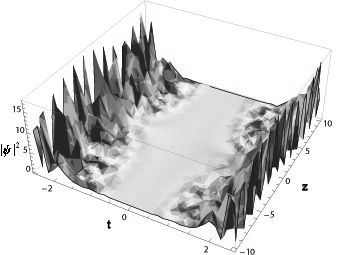}}
\hspace{0.35in} \subfigure[Dark soliton train, $m=0.5$,
$\tau_0=0.375$]{
\includegraphics[scale=0.4]{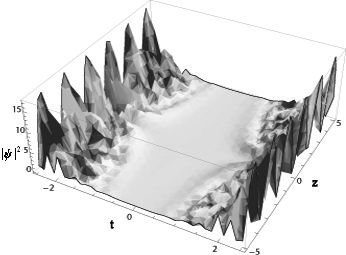}}\\
\vspace{0.5in} \subfigure[Bright soliton, $m=1$, $\tau_0=1$]
{\includegraphics[scale=0.45]{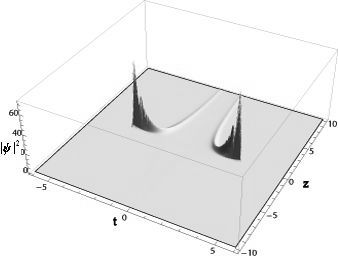}}
\hspace{0.35in} \subfigure[Dark soliton, $m=1$, $\tau_0=1$ ]{
\includegraphics[scale=0.45]{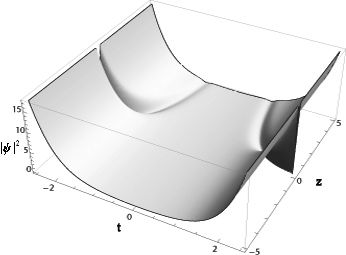}}\\
\vspace{0.5in} \subfigure[$\gamma(t)$ versus $t$, $\gamma_0 = -0.5$
]{\includegraphics[scale=0.45]{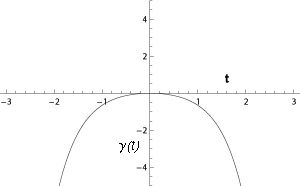}}
 \subfigure[$\gamma(t)$ versus $t$, $\gamma_0 = 0.5$
]{\includegraphics[scale=0.45]{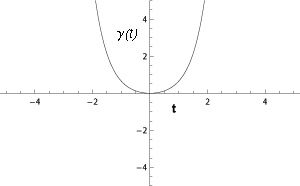}}
\caption{$c(t)=coth(t)$ in the expulsive oscillator regime. Plots of
the bright [(a),(c)] and dark [(b),(d)] soliton solutions,
$\psi(z,t)=\sqrt{A(t)}F\{(A(t)[z-l(t)])/\tau_0,m\}\exp[\imath
a(t)-\frac{\imath}{2}c(t)z^2]$ with $F=\cn(T/\tau_0,m)$ and
$F=sn(T/\tau_0,m)$ respectively, are shown. The plots (a)-(d) show
$|\psi(z,t)|^2$ versus $t$ and $z$  and the plots (e) and (f) show
the variation of the nonlinearity $\gamma(t)$ with $t$. The values
of the constant parameters are $A_0=0.5$, $l_0 =5$, $t_0=0.9$ and
$B=2$.}
\end{figure*}


\begin{figure*}
\centering \subfigure[Bright soliton train, $m=0.5$, $\tau_0=0.375$]
{\includegraphics[scale=0.5]{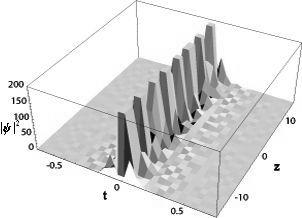}}
\hspace{0.35in} \subfigure[Dark soliton train, $m=0.5$,
$\tau_0=0.375$]{
\includegraphics[scale=0.5]{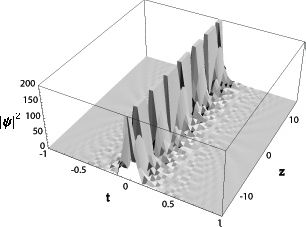}}\\
\vspace{0.5in} \subfigure[Bright soliton, $m=1$, $\tau_0=1$]
{\includegraphics[scale=0.45]{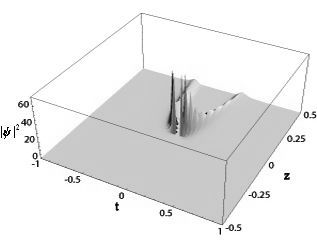}}
\hspace{0.35in} \subfigure[Dark soliton, $m=1$, $\tau_0=1$ ]{
\includegraphics[scale=0.45]{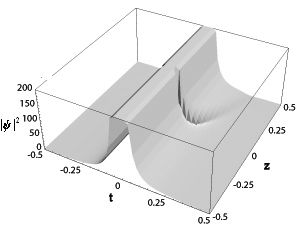}}\\
\vspace{0.5in} \subfigure[$\gamma(t)$ versus $t$, $\gamma_0 = -0.5$
]{\includegraphics[scale=0.5]{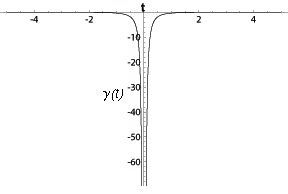}}
 \subfigure[$\gamma(t)$ versus $t$, $\gamma_0 = 0.5$
]{\includegraphics[scale=0.5]{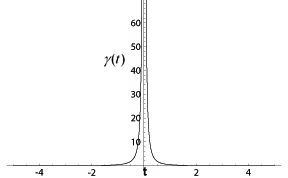}}
\caption{$c(t)=-coth(t)$ in the expulsive oscillator regime. Plots
of the bright [(a),(c)] and dark [(b),(d)] soliton solutions,
$\psi(z,t)=\sqrt{A(t)}F\{(A(t)[z-l(t)])/\tau_0,m\}\exp[\imath
a(t)-\frac{\imath}{2}c(t)z^2]$ with $F=\cn(T/\tau_0,m)$ and
$F=sn(T/\tau_0,m)$ respectively, are shown. The plots (a)-(d) show
$|\psi(z,t)|^2$ versus $t$ and $z$  and the plots (e) and (f) show
the variation of the nonlinearity $\gamma(t)$ with $t$. The values
of the constant parameters are $A_0=0.5$, $l_0 =5$, $t_0=0.8812$ and
$B=2$.}
\end{figure*}


\begin{figure*}
\centering \subfigure[Bright soliton train, $m=0.5$, $\tau_0=0.375$]
{\includegraphics[scale=0.4]{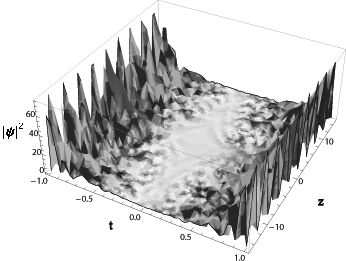}}
\hspace{0.35in} \subfigure[Dark soliton train, $m=0.5$,
$\tau_0=0.375$]{
\includegraphics[scale=0.4]{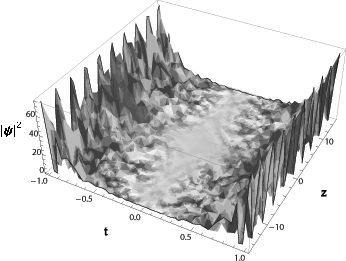}}\\
\vspace{0.5in} \subfigure[Bright soliton, $m=1$, $\tau_0=1$]
{\includegraphics[scale=0.4]{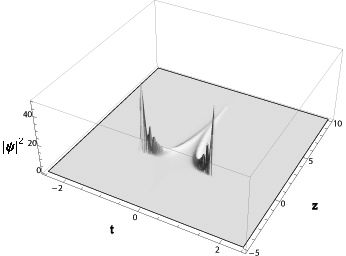}}
\hspace{0.35in} \subfigure[Dark soliton, $m=1$, $\tau_0=1$ ]
{\includegraphics[scale=0.4]{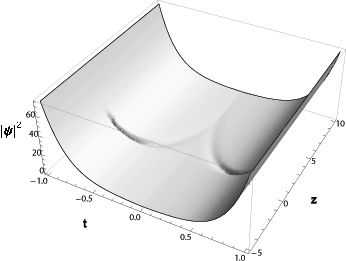}}\\
\vspace{0.5in} \subfigure[$\gamma(t)$ versus $t$, $\gamma_0 = -0.5$
]{\includegraphics[scale=0.45]{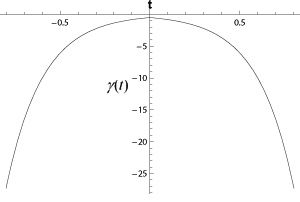}}
\hspace{0.35in} \subfigure[$\gamma(t)$ versus $t$, $\gamma_0 = 0.5$]
{\includegraphics[scale=0.4]{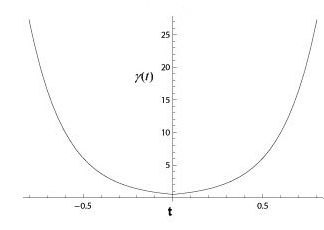}}
\caption{Attractive $\delta$ potential in the expulsive oscillator
regime, $\phi(t)=\sqrt{V_0/2}\exp(-V_0|t|/2)$. Plots of the bright
[(a),(c)] and dark [(b),(d)] soliton solutions,
$\psi(z,t)=\sqrt{A(t)}F\{(A(t)[z-l(t)])/\tau_0,m\}\exp[\imath
a(t)-\frac{\imath}{2}c(t)z^2]$ with $F=\cn(T/\tau_0,m)$ and
$F=sn(T/\tau_0,m)$ respectively, are shown. The plots (a)-(d) show
$|\psi(z,t)|^2$ versus $t$ and $z$  and the plots (e) and (f) show
the variation of the nonlinearity $\gamma(t)$ with $t$. The values
of the constant parameters are $V_0=10$, $A_0=0.5$, $l_0=5$.}
\end{figure*}


\begin{figure*}
\centering \subfigure[Bright soliton train, $m=0.5$, $\tau_0=0.375$]
{\includegraphics[scale=0.4]{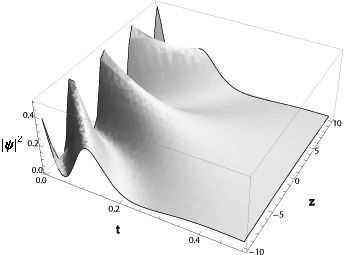}}
\hspace{0.35in} \subfigure[Dark soliton train, $m=0.5$,
$\tau_0=0.375$]{
\includegraphics[scale=0.4]{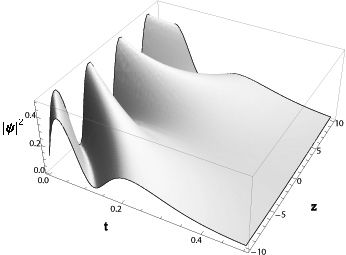}}\\
\vspace{0.5in} \subfigure[Bright soliton, $m=1$, $\tau_0=1$]
{\includegraphics[scale=0.4]{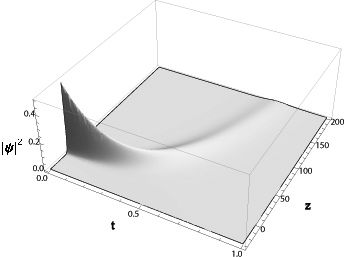}}
\hspace{0.35in} \subfigure[Dark soliton, $m=1$, $\tau_0=1$ ]{
\includegraphics[scale=0.4]{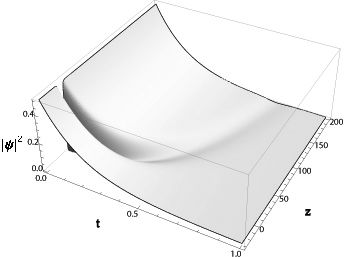}}\\
\vspace{0.5in} \subfigure[$\gamma(t)$ versus $t$, $\gamma_0 = -0.5$
]{\includegraphics[scale=0.45]{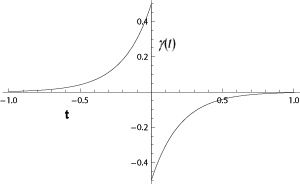}}
\hspace{0.35in} \subfigure[$\gamma(t)$ versus $t$, $\gamma_0 = 0.5$
]{\includegraphics[scale=0.45]{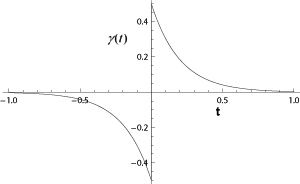}}
\caption{Attractive $\delta$ potential in the expulsive oscillator
regime, $\phi(t)=\sqrt{2/V_0}\exp(V_0 t/2)$. Plots of the bright
[(a),(c)] and dark [(b),(d)] soliton solutions,
$\psi(z,t)=\sqrt{A(t)}F\{(A(t)[z-l(t)])/\tau_0,m\}\exp[\imath
a(t)-\frac{\imath}{2}c(t)z^2]$ with $F=\cn(T/\tau_0,m)$ and
$F=sn(T/\tau_0,m)$ respectively, are shown. The plots (a)-(d) show
$|\psi(z,t)|^2$ versus $t$ and $z$  and the plots (e) and (f) show
the variation of the nonlinearity $\gamma(t)$ with $t$. The values
of the constant parameters are $V_0=10$, $A_0=0.5$, $l_0 =5$.}
\end{figure*}
\end{document}